\documentclass[useAMS,usenatbib]{mn2e}
\usepackage{graphicx}
\usepackage[figuresright]{rotating}
\usepackage{txfonts}
\newcommand{\ii}{\'\i}
\newcommand\ion[2]{#1$\;${\scshape{#2}}}

\title[Field galaxies  at intermediate redshifts 
($0.2 < \emph{z} < 0.8$) in the direction of the galaxy 
cluster LCDCS-S001]{Field galaxies at intermediate redshift ($0.2 < \emph{z} < 0.8$) 
in the direction of the galaxy cluster LCDCS-S001}

\author[A.~C.~Krabbe, S.~B.~Rembold and M.~G.~Pastoriza]{
A.~C.~Krabbe\thanks{E-mail:
angela.krabbe@ufrgs.br}, S.~B.~Rembold and M.~G.~Pastoriza
\\
Instituto de F\ii sica, Universidade Federal 
do Rio Grande do Sul, Av.~ Bento Gonçalves,9500, 
Cep 91359-050, Porto Alegre, RS, Brazil}
 
\begin{document}

\date{Accepted -. Received -.}

\pagerange{\pageref{firstpage}--\pageref{lastpage}} \pubyear{2006}

\maketitle

\label{firstpage}

\begin{abstract}
We present spectroscopic and photometric analysis for eight field galaxies in the direction
of the galaxy cluster LCDCS-S001. 
The spectra were obtained with the GMOS instrument in the Gemini 
South Observatory. The objects were selected in an $i'$ band image and 
the multi-object spectroscopic  observations were centered at 7\,500 \AA. For the galaxies 
ID\,440 and ID\,461 we have determined redshifts of
$\emph{z}=0.7464$ and $\emph{z}=0.7465$, respectively. 
For the other six galaxies we have 
confirmed the redshift calculated by \citet{rembold06}. 
The redshifts of the field galaxies are in the range of $0.2201 < \emph{z} < 0.7784$.
We determined the blue and visual luminosities and they are brighter than 
M$_{B}=-18.64$. The galaxies ID\,180, ID\,266, ID\, 461 follow the Faber-Jackson relation of the 
Coma and Virgo early-type galaxies, and therefore do not present 
a brightening of the $B$ luminosity as observed in galaxies at higher redshifts.
The stellar velocity dispersion was measured 
for five galaxies (ID\,146, ID\,180, ID\,266, ID\,428 and ID\,440) and estimated
 to be in the range  of
$200 < \sigma <  346\, \rm{ km\, s^{-1}}$. 
Lick indices were measured and used 
to determine the stellar population properties of galaxies ID\,120 and ID\,146, 
by means of  spectral  synthesis. The first galaxy, ID\,120, presents in its
spectrum absorption and emission lines, and we have found that the 
main contribution in the flux at $\lambda$ 5870 \AA\,  is 
of a  0.1 Gyr stellar population of solar
metallicity. For ID\,146,  the  dominant flux  contribution 
at $\lambda$ 4200 \AA\, is provided by  a stellar population  of 10 Gyr of 
subsolar metallicity. From stellar population synthesis we
estimated reddening values of   
$E(B-V)=0.90$  and $E(B-V)=0.82$ for ID\,120 and ID\,146, respectively. 
According to classical diagnostic diagrams the emission lines present in the 
spectrum of ID\,120 indicate that it is a starburst galaxy.

\end{abstract}


\section{Introduction}
Galaxy clusters are one of the fundamental sites to study the formation
and evolution of galaxies. Many observational studies in cluster galaxies have 
revealed that galaxy properties such as colours, morphology and stellar populations
vary at different redshifts. \citet{butcher84} have found 
in intermediate redshift clusters
that the fraction of  blue galaxies  \citep{butcher84} and galaxies with star formation  
\citep{poggianti99} is higher than in  clusters of the local universe. 
Luminosity evolution was also observed for galaxies at intermediate redshift with respect to
the local galaxies. There was a brightening in blue
magnitude in the Faber-Jackson relation \citep{ziegler97,fritz05}.

Differences between the stellar population of  cluster and field galaxies  
at intermediate and higher redshifts 
have been revealed. \citet{serote05} have found that the field galaxies seem to host less 
evolved stellar populations than 
cluster members; and  
\citet{vandokkum01}  found that the field early-type galaxies are about 20 \% younger than 
cluster early-type galaxies. Moreover, significant offsets between field 
and cluster galaxies were derived at \emph{z}$=0.7$ by \citet{treu02}. However,
few  observational works on clusters and field galaxies at intermediate 
redshifts  have been published to confirm these results.

In a previous work \citet{rembold06} (hereafter Paper I) presented a 
study on the kinematic parameters and stellar population properties of galaxies of 
cluster  LCDCS-S001 ($\emph{z}=0.7$). In the field of view of this galaxy cluster 
 several galaxies were  observed at intermediate redshifts which are not cluster
members. In particular,  ID\,120 at $\emph{z}=0.2201$ shows in its spectrum
very bright emission lines apparently typical of starburst galaxies.  
We have studied these galaxies and  report 
in the present paper  the more important  spectroscopic and photometric results such 
as redshifts, stellar population, 
blue luminosity and for ID\,120 we discuss 
the properties of the
ionized gas. The paper is structured as follows. In Sect. 2  we describe 
 the observations and data reduction. In Sect. 3 we present the
determination of redshifts and velocity dispersions. In Sect. 4  we describe the determination
of absolute magnitudes and the relation between luminosity and velocity 
dispersion. Sec. 5 deals with the stellar population synthesis, describing the 
measurements of Lick indices, the method used and the results obtained.
In Sec. 6 we determine the properties of the nebular gas and the nature of 
the ionization source. 
The conclusions are given in Sect. 7.
Throughout this paper we adopt  a cosmological model  with 
$H_0=71 \rm{ km\, s^{-1}} $, $\Omega_M=0.27$ and $\Omega_V=0.73$.


\section{Observations and data reductions}
The current paper is based on spectroscopic data obtained on 
March 2004 with the 
Gemini Multi-Object Spectrograph (GMOS)
of the Gemini South Observatory. The GMOS mask was built using as reference an 
i' filter image obtained with GMOS,
which covered an area of $330\times330$\,arcsec$^2$, with a total
integration time of 300 seconds. The detector was binned at $2\times 2$
pixels, giving a spatial scale of 0.146\,arcsec\,pxl$^{-1}$. 
The objects included in this study are listed in Table \ref{redshift}. We used a grid 
of 400 grooves mm$^{-1}$. The spectra were obtained in the 5\,560 to 9\,720  \AA\, range with 
a dispersion of 0.67 \AA\,pxl$^{-1}$
and a spectral resolution  of $R\simeq1918$. Two masks were needed to 
include all observed objects.
The exposure time for each mask was limited to 1200 seconds 
to minimize the effects of cosmic rays and to obtain a signal-to-noise higher than 3.

\begin{figure}
\centering
\includegraphics*[width=\columnwidth]{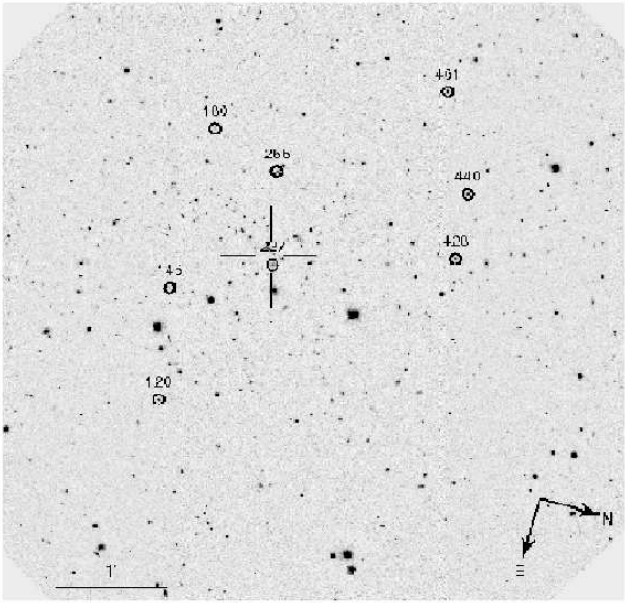}					     
\caption{$i'$ band image of the galaxy cluster LCDCS-S001 
with the field galaxies marked with circles and the visual center of the 
cluster marked with a cross.}
\label{field}
\end{figure}

\begin{figure*}
\centering
\begin{tabular}{cccc}
ID\,120 \, ( \emph{z}= 0.2201 )& ID\,146\, ( \emph{z}= 0.3910 ) & ID\,180 \, ( \emph{z}= 0.6508 ) & ID\,237 \, ( \emph{z}= 0.2607 )\\
{\includegraphics[width=4cm]{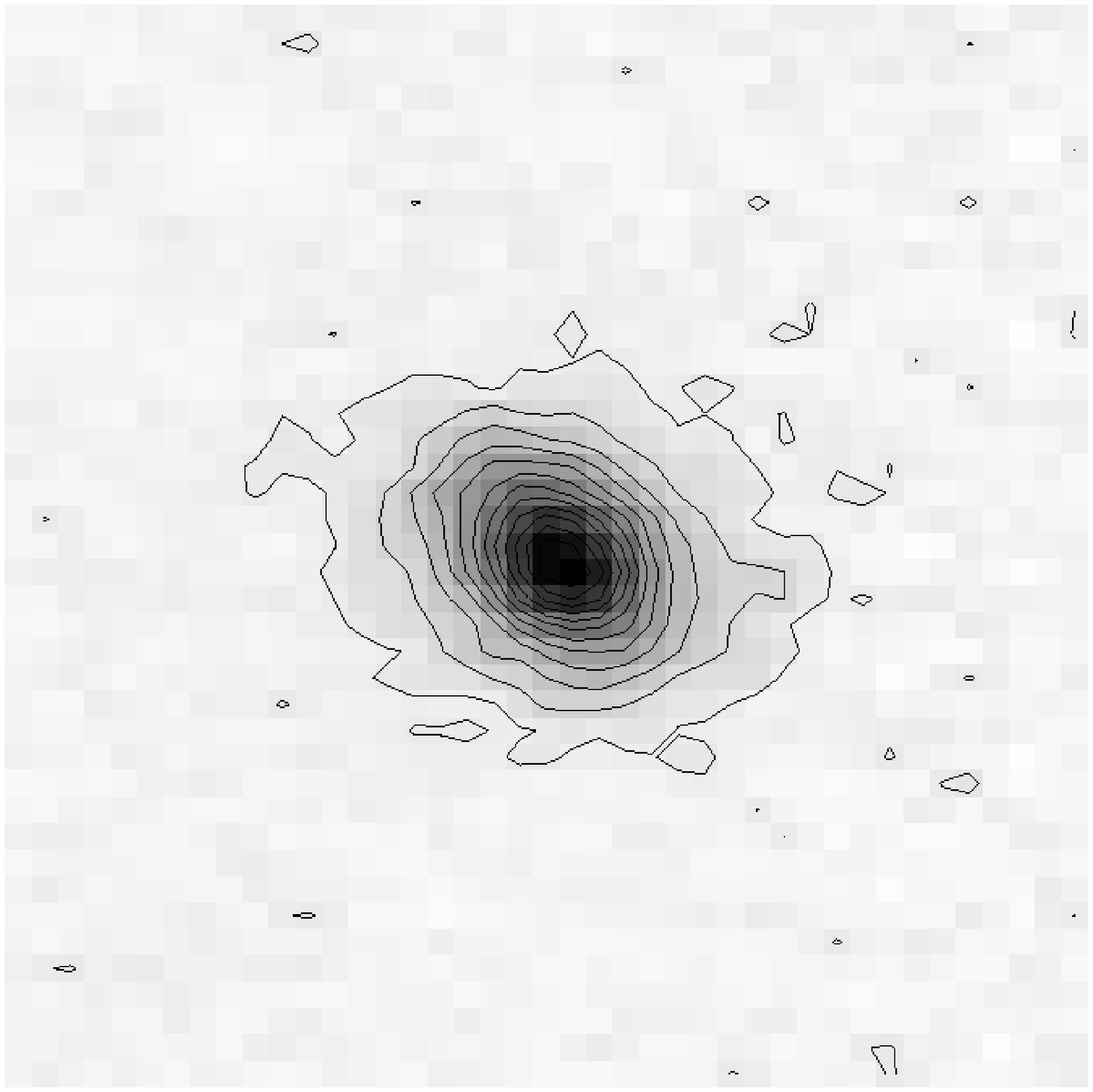}} &
{\includegraphics[width=4cm]{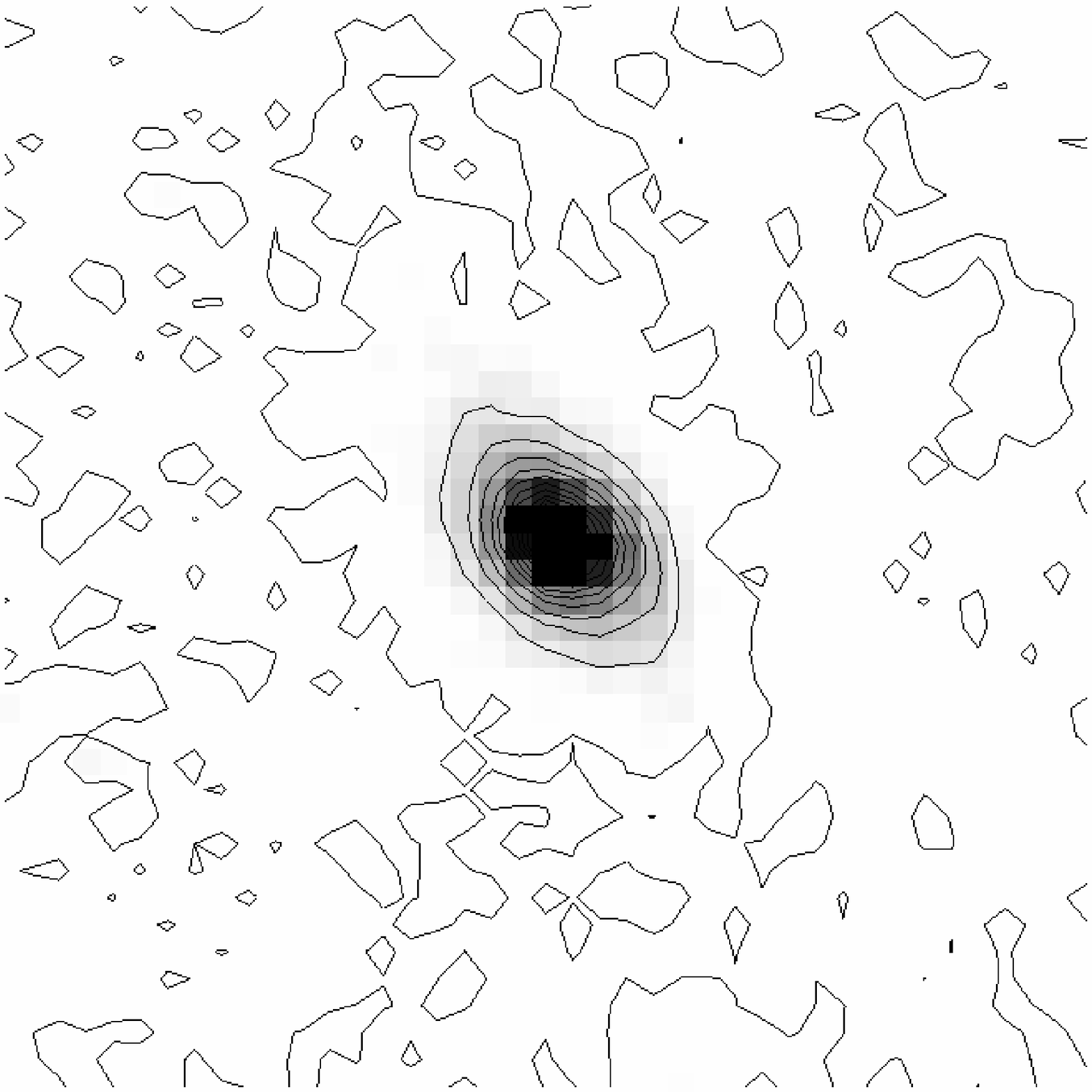}} 
&
{\includegraphics[width=4cm]{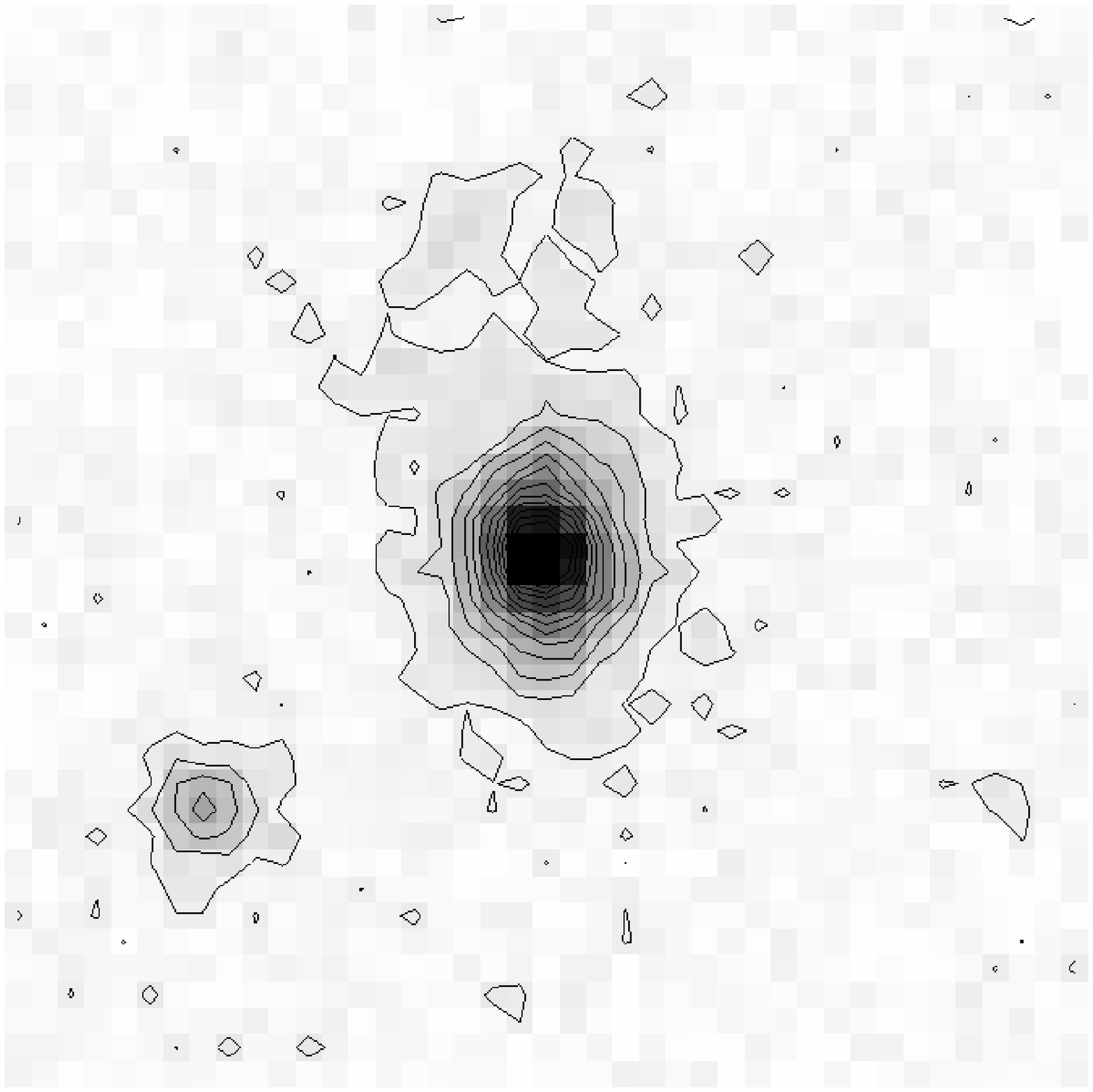}} &
{\includegraphics[width=4cm]{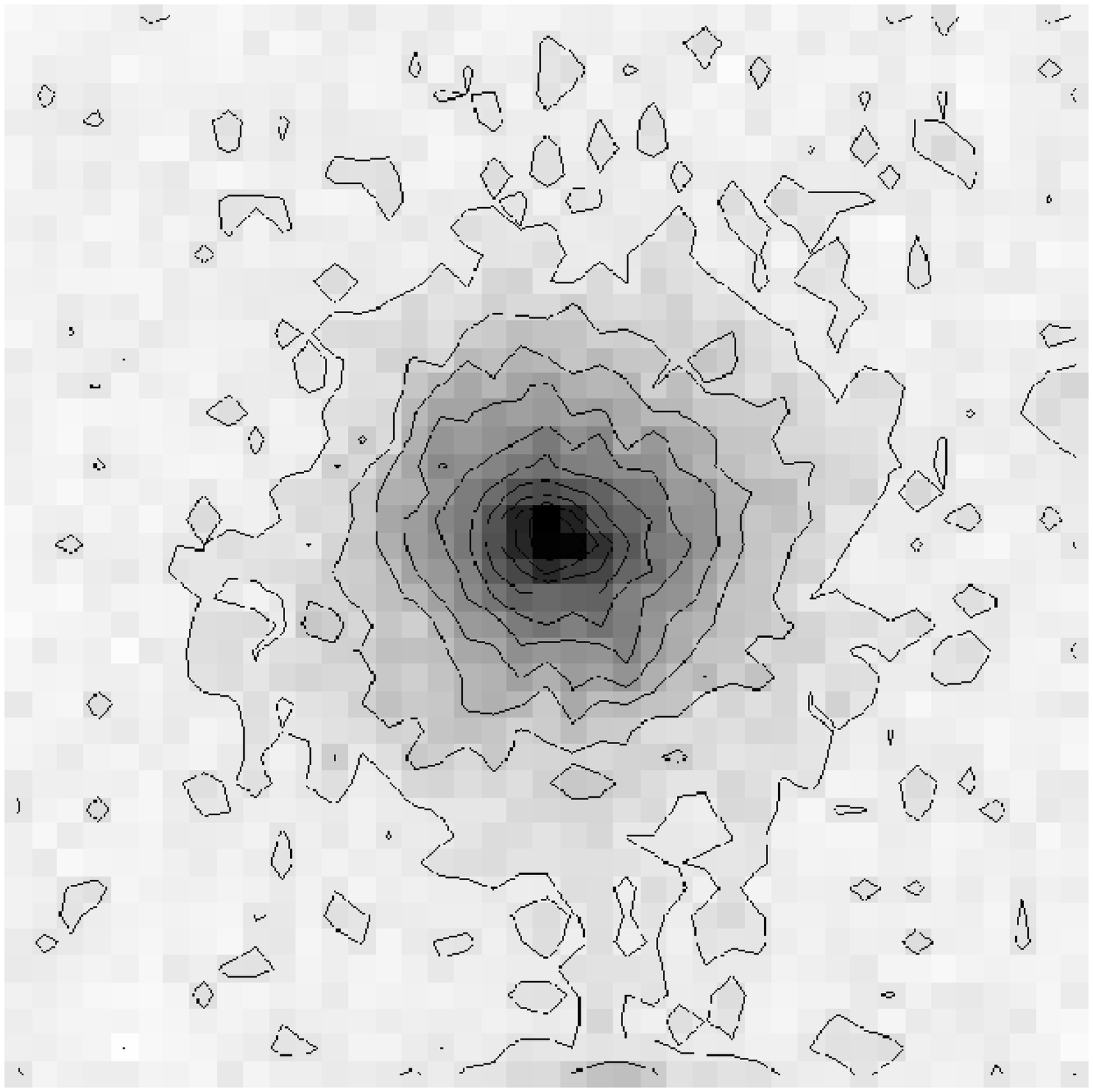}} \\
ID\,266 \, ( \emph{z}= 0.7784 )& ID\,428 \, ( \emph{z}= 0.7784 )& ID\,440 \, ( \emph{z}= 0.7464 )& ID\,461 \, ( \emph{z}= 0.7465 )  \\
{\includegraphics[width=4cm]{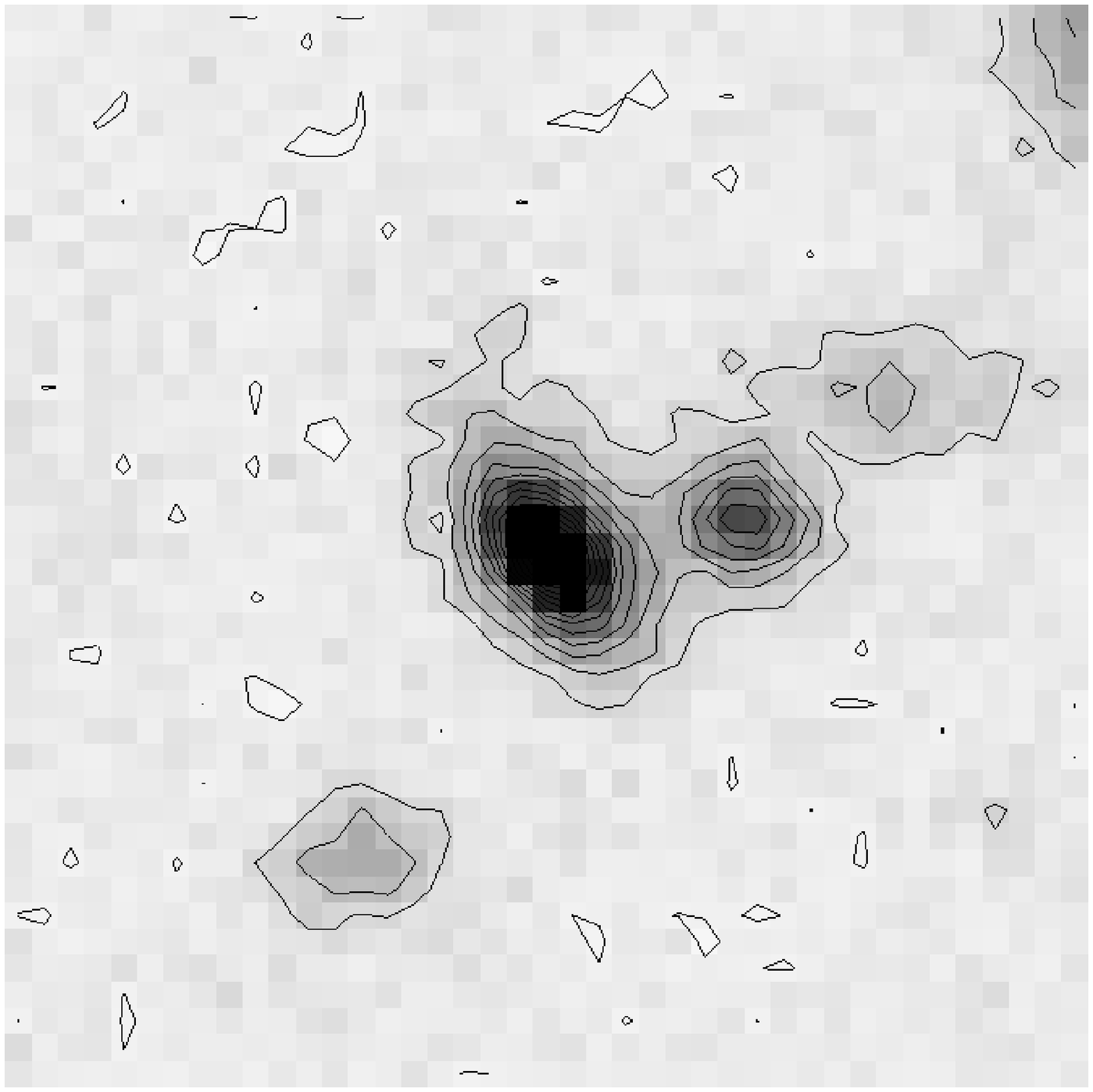}} &
{\includegraphics[width=4cm]{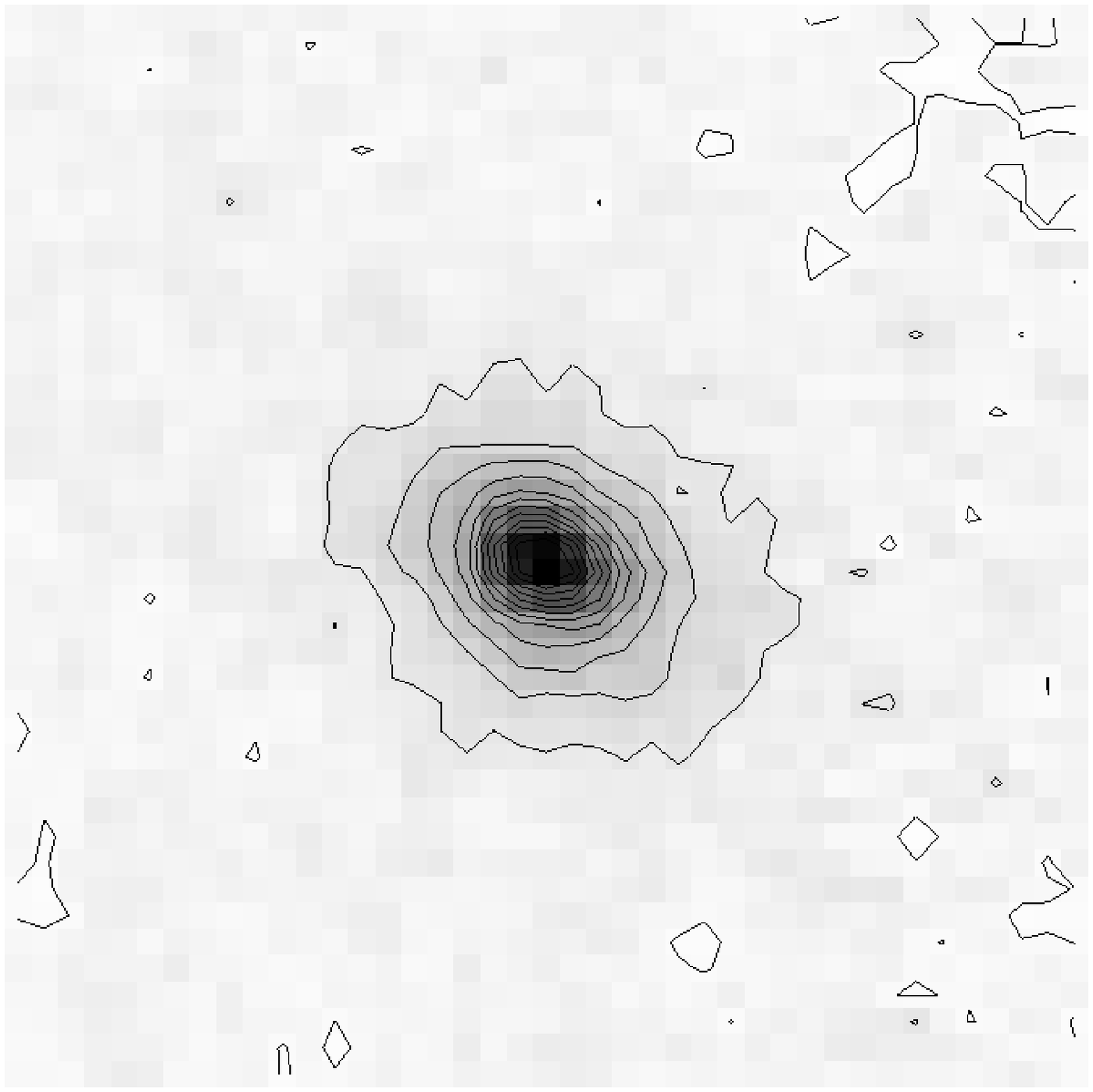}} &

{\includegraphics[width=4cm]{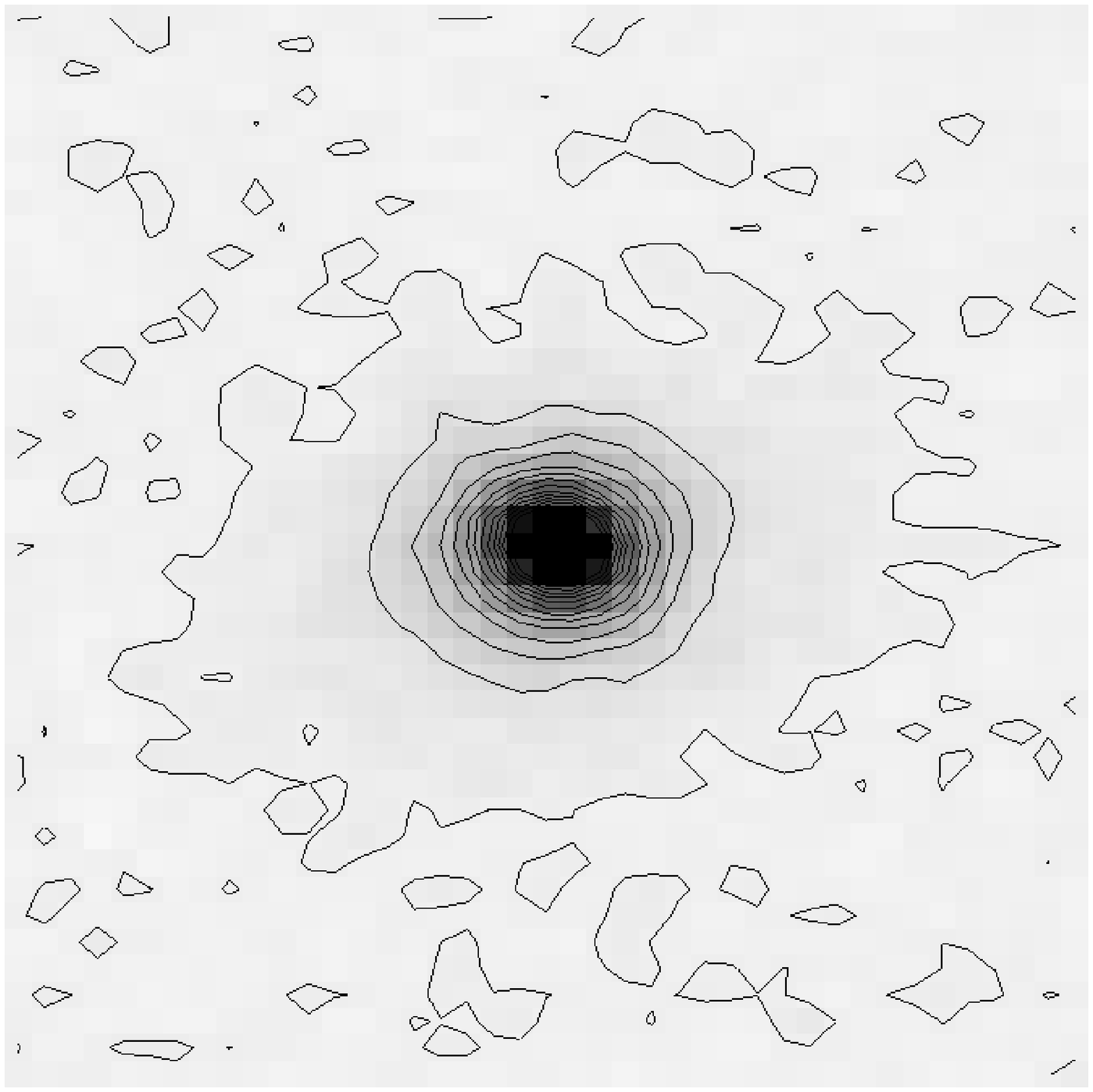}} &
{\includegraphics[width=4cm]{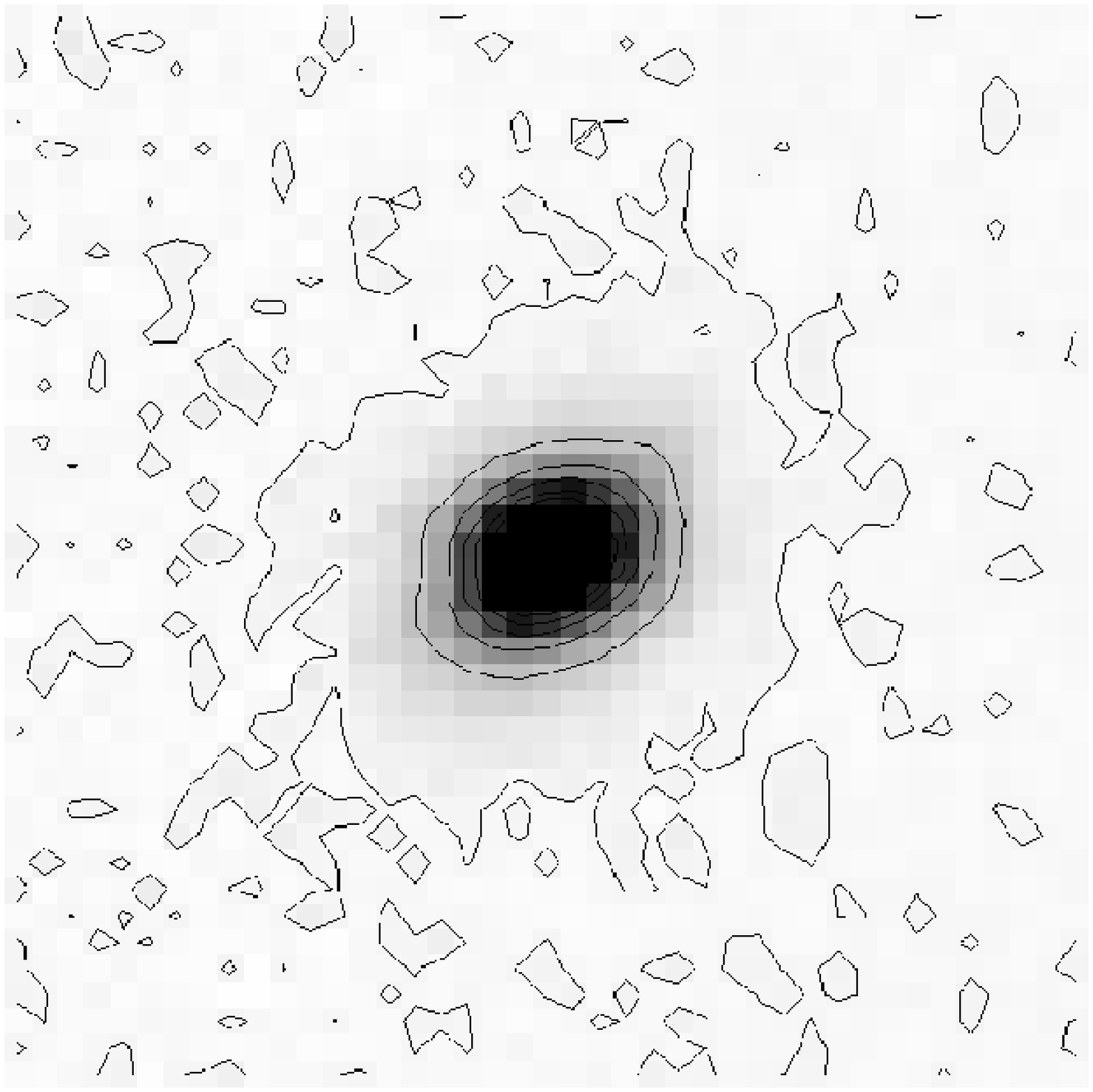}} \\
\end{tabular}
\caption{$i'$ band images and contours of the observed galaxies. All 
frames are 5.99$\times$5.99 arcsec$^2$.}
\label{contour}
\end{figure*}

The data reduction followed the standard procedures of 
bias correction, flat-fielding, cosmic ray cleaning, 
sky subtraction, wavelength and flux calibrations, and  1D spectra extraction, 
and was made using 
mainly the $IRAF$ software. The procedures we followed in the data reduction are described 
in detail in   Paper I. Figure \ref{field} shows the full area imaged by GMOS and the analyzed objects,
which are numbered and marked with circles. 
Figure \ref{contour} shows the isophotal maps in the $i'$ band for each galaxy of 
our sample. In Figure 3 we  present one dimensional spectra where the most 
prominent lines are identified and 
listed in Table \ref{redshift}.

\begin{table*}
\caption{Coordinates, redshifts, absolute magnitudes and spectral lines of the observed galaxies}
\label{redshift}
\begin{tabular}{lccccccccl}
\noalign{\smallskip}
\hline
\hline
\noalign{\smallskip}
ID    & $\alpha$(2000) & $\delta$(2000) &\emph{z} & $\sigma$ (km s$^{-1})$ &$i'_{AB}$&  M$_{AB,i'}$&  M$_{B}$& M$_{V}$ &Spectral lines  \\
\noalign{\smallskip}
\hline
\noalign{\smallskip}

120   & $10^{\rm{h}}06^{\rm{m}}25\fs20$ & $-12^{\rm{h}}58^{\rm{m}}56\fs6$ &  0.2201& ...&21.42
 &-18.75 & ... &-18.20& CH $\lambda 4300$, $\rm H\beta$ $\lambda 4861$, [\ion{O}{iii}] $\lambda 4959$, 
[\ion{O}{iii}] $\lambda 5007$, \\ 
 & &  & & &&&&&\ion{Mg}{i} $\lambda 5167$, NaD $\lambda 5890$,[\ion{N}{ii}] $\lambda 6548$,
$\rm H\alpha$ $\lambda 6563$,
\\
& &  & & &&&&&[\ion{N}{ii}] $\lambda 6584$,[\ion{S}{ii}] $\lambda 6717$, [\ion{S}{ii}]$ \lambda 6731$ \\

146   & 10 ~06 ~21.10 & -12 ~59 ~07.7 &  0.3910&217$\pm$19 & 21.50
&-20.11&-18.64&...& \ion{Fe}{i} $\lambda 3631$, \ion{Fe}{i} $\lambda 3720$, 
 \ion{Fe}{i} $\lambda 3827$, $\rm H9$ $\lambda 3835$, \\
& &  & & &&&&& \ion{Ca}{ii} $\lambda 3934$,\ion{Ca}{ii} $\lambda 3968$, $\rm H\delta$ $\lambda 4101$,
\ion{Ca}{i} $\lambda 4226$,\\
& &  & & &&&&& CH $\lambda 4300$,\ion{Fe}{i} $\lambda 4383$,$\rm H\beta$ $\lambda 4861$, 
 [\ion{N}{i}] $\lambda 5200$\\

180   & 10 ~06 ~14.90 & -12 ~59 ~07.9 &  0.6508 & 274$\pm$58 & 21.75 
& -21.20&-20.66 &...& \ion{Ca}{ii} $\lambda 3934$, \ion{Ca}{ii} $\lambda 3968$, 
CH $\lambda 4300$, $\rm H\gamma$ $\lambda 4340$	 \\

237   & 10 ~06 ~19.22 & -12 ~58 ~17.0 &  0.2607& ...&20.82 
&-19.46 &... &-18.93& $\rm H\beta$ $\lambda 4861$, [\ion{O}{iii}] $\lambda 5007$\\

266   & 10 ~06 ~15.80 & -12 ~58 ~28.8 &  0.7784& 200$\pm$23& 21.89
&-21.54 &-21.29 &...& $\rm H10$ $\lambda 3798$, $\rm H9$ $\lambda 3835$, $\rm H8$ $\lambda 3889$, 
\ion{Ca}{ii} $\lambda 3934$,   \\ 
& &  & & &&&&& \ion{Ca}{ii} $\lambda 3968$  \\

428   & 10 ~06 ~17.14 & -12 ~56 ~41.6 &  0.7784& 278$\pm$89&21.82 
&-22.08 &-22.45 &...&  \ion{Ca}{ii} $\lambda 3934$, \ion{Ca}{ii} $\lambda 3968$  \\   

440   & 10 ~06 ~14.71 & -12 ~56 ~44.8 &  0.7464&346$\pm$12 &20.98 
&-22.34 &-22.01 &...& \ion{Ca}{ii} $\lambda 3934$, \ion{Ca}{ii} $\lambda 3968$  \\    
  
461   & 10 ~06 ~11.24 & -12 ~57 ~10.8 &  0.7465&...  & 20.47 
& -22.85&-22.52 &...& 
 \ion{Ca}{ii} $\lambda 3934$, \ion{Ca}{ii} $\lambda 3968$   \\
\noalign{\smallskip}
\hline
\noalign{\smallskip}
\end{tabular}\\
\begin{minipage}[c]{18.0cm}
{\it Conventions:} $\alpha$, $\delta$: equatorial coordinates;
\end{minipage}
\end{table*}

\begin{figure*}
\label{all_spectra}
\centering
\includegraphics*[height=14cm]{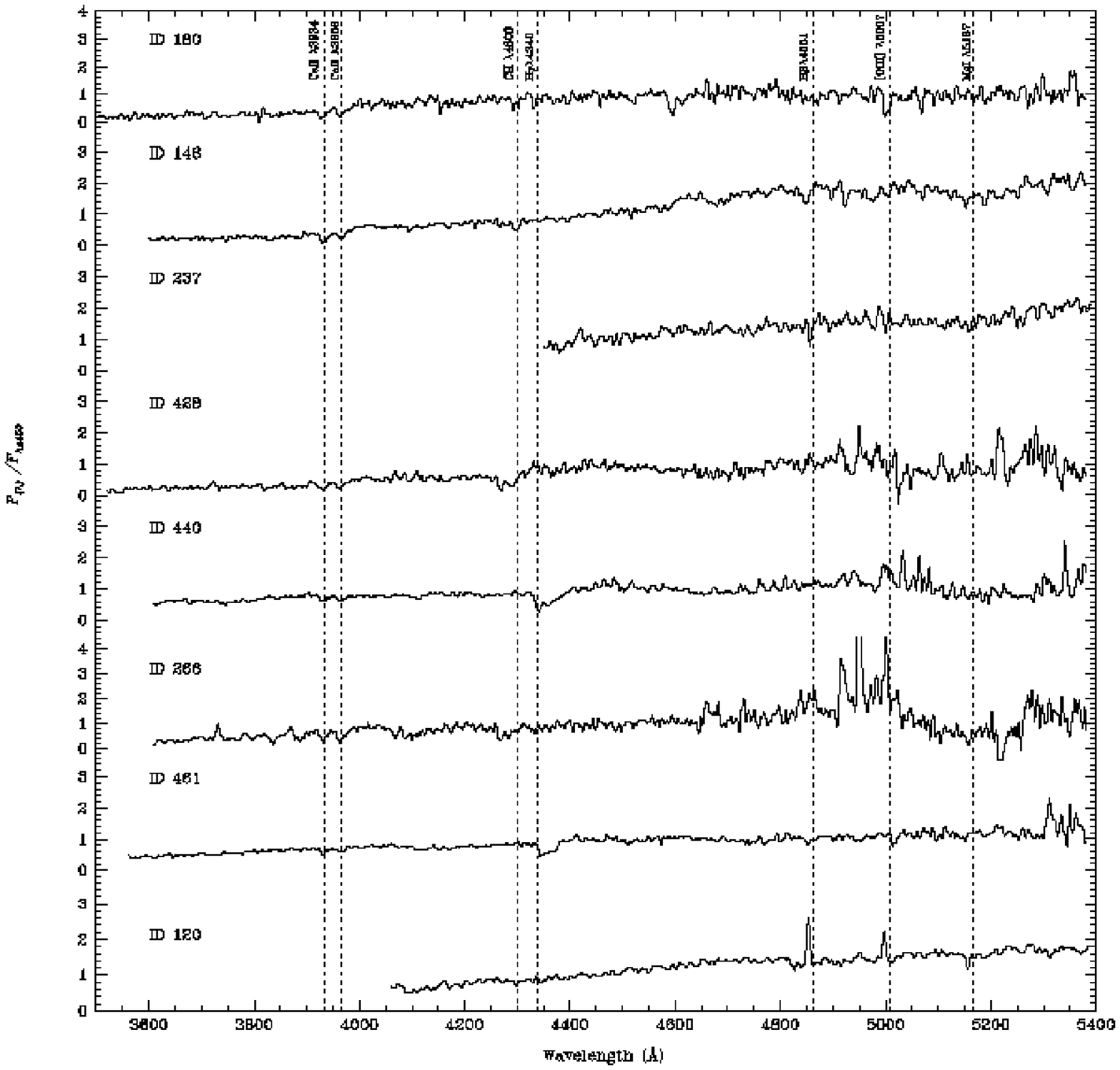}						     
\caption{Spectra of the field galaxies in the of 3500 to 5400 \AA. The most prominent 
absorption and emission lines are marked.}
\end{figure*}

\section{Determination of redshifts and velocity dispersions}
\label{dispersion}
The spectroscopic redshifts for ID\,120, ID\,146, ID\,180, ID\,237, ID\,266, and ID\,428
were taken from  Paper I. In this work we determined the redshift only for  
the galaxies  ID\,440 and ID\,461.  The redshift of these galaxies was 
calculated by the average recession velocity given by the 
individual lines and estimated to be around   $\emph{z}=0.7464$ and $\emph{z}=0.7465$ for 
ID\,440 and ID\,461, respectively. The galaxy cluster LCDCS-S001 has a 
median cluster redshift of 0.709 and an estimated velocity
dispersion of $\sigma = 727\, \rm km s^{-1}$ (see Paper I), therefore the last 
galaxies are not cluster  members. 
Note that the galaxies of our sample  can be separated into two sets: 
one (ID\,120, ID\,146, ID\,237)  located to the east of the cluster center, 
in the redshift range $0.2201< \emph{z} <0.3910$;
and the other set of galaxies (ID\,266, ID\,428, ID\,440, ID\,461) is 
to the west of the center and the redshift is in the range $0.7464 < \emph{z} < 0.7784$. 
The galaxy ID\,180 
to the west of the cluster center is at redshift $\emph{z}=0.6508$. 
The error associated with the redshift for each galaxy, estimated from the standard 
deviation of the individual redshifts, are practically negligible, below  1\%. 
 
The velocity dispersions, $\sigma$, of the observed galaxies were determined from  the measurements 
of corrected full-width-at-half-maximum, $FWHM$, assuming Gaussian profiles for the absorption lines. 
The  effects of 
the instrumental dispersion and resolution on the observed FWHM 
were estimated from  the measurements of 
the absorption lines \ion{Ca}{ii}$\lambda\,8542$ and \ion{Ca}{ii}$\lambda\,8662$ 
observed in the spectrum of a G star 
identified in the field.  We have found that the velocity dispersions for most of our  
galaxies (ID\,146, ID\,180, ID\,266, ID\,428, ID\,440)  are in the range 
$200 < \sigma < 346\, \rm{km\, s^{-1}}$. 
The galaxies ID\,120, ID\,237 and ID\,461 have velocity dispersions of the order of the
instrumental resolution and therefore are not estimated. 
The values of redshift and velocity dispersions $\sigma$ are listed in Table \ref{redshift}.

\section{Determination of absolute magnitudes}
In this paper we have estimated the absolute  magnitudes M$_{B}$ and M$_{V}$ for
our sample of galaxies using  the values of $i'_{AB}$ photometric  
magnitude  taken from Paper I and the $B$ and $V$ magnitudes derived from  the observed spectra.  
In order to calculate these  magnitudes we have calibrated the  observed spectra     
using the $i'_{AB}$ photometric magnitude.
Note that the GMOS filters are defined in the $AB$ system as 

\begin{equation}
i'_{AB} =-2.5\,\log_{10} { f_\nu}{\rm ( erg\,s^{-1}\,cm^{-2}\,Hz^{-1})- 48.60}.
\end{equation}

We  have transformed the spectra  to the restframe system and applied for the fluxes 
the $(1+z)^2$ correction for cosmological dimming. Then, we have calculated the
$B$ and $V$ magnitudes  from the  spectra by integrating the fluxes  through a set of 
filter response curves using the  {\it sbands} routine of the {\it 
onedspec} package of the {\it NOAO/IRAF}. Finally, we 
added this magnitude to the distance modulus, calculated
as a function of the galaxy redshift \emph{z} adopting    
the cosmological model assumed in this paper. 
The values of M$_{B}$, M$_{V}$ and also of M$i'_{AB}$ (taken from Paper I) 
are listed in Table \ref{redshift}. We verify that our field galaxies  
display a correlation between redshift and absolute magnitude in $i'_{AB}$ band 
as   well as in the $V$ band. This means that high luminosity  galaxies are detected at higher 
redshift, while low luminosity galaxies are detected mainly at lower redshifts. This
behavior is a selection effect for samples that are chosen based on magnitudes. 
The total error associated with the absolute magnitudes was estimated to be in the range
of  0.12 to 0.37 mag. This error was estimated considering (a) the uncertainties due  to  
the continuum signal-to-noise ratio in each spectrum and (b) the uncertainties of the
$i'$ magnitude, which are mainly due to errors on the count fluctuations, the 
assumption of the zero-point and the lack of a 
color term correction. 

The field galaxies in our sample are  very luminous with  blue luminosity in the range of 
$-18.64 < {\rm M}_{B} < -22.52$. Their luminosities are comparable with the luminosities of galaxies at intermediate redshifts 
\citep{mouhcine06,nakamura06,ziegler97}. 

\section{Faber-Jacskon relation}
It is  well known that elliptical and lenticular galaxies  follow the scaling relation 
between galaxy  luminosity and velocity dispersion \citep{faber76}, which is widely used 
to obtain important information about elliptical galaxies. This is called the Faber-Jackson 
relation and is valid both for local and intermediate redshift galaxies.
We searched in our sample if there are early-type 
galaxies that follow this relation 
and present  a brightening of blue
magnitude with respect to local clusters of galaxies as inferred by \citet{ziegler97} .

\begin{figure}
\centering
\includegraphics*[width=\columnwidth]{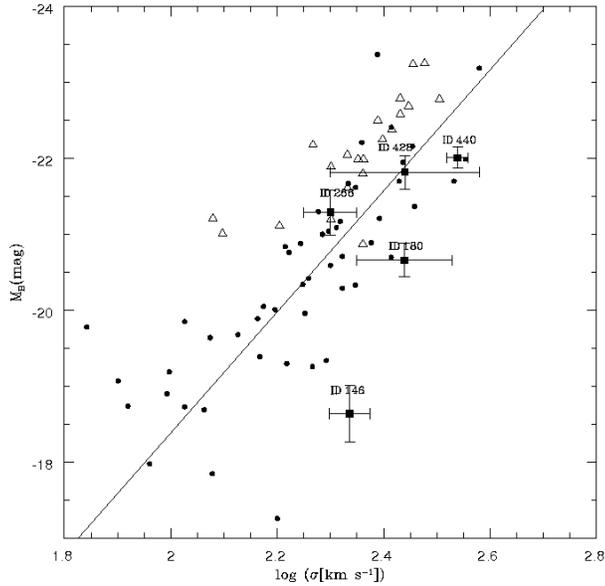}					     
\caption{Faber-Jackson relation in absolute $B$ magnitude for 5 galaxies 
in the field of the galaxy cluster LCDCS-S001(square with error bars). The circles represent  
the Coma and Virgo sample of  \citet{dressler87} and the triangles represent 
the elliptical galaxy sample at $\emph{z} =0.37$ of \citet{ziegler97}. The solid
line is the best fit for the Coma  sample, as derived by  \citet{ziegler97}.}
\label{faber}
\end{figure}

In Figure \ref{faber} the M$_{B}$
versus $\sigma$ relation for the galaxies  ID\,146, ID\,180, ID\,266, ID\,428 and  ID\,440 is plotted and also compared with
data of a sample of Coma and Virgo elliptical galaxies  \citep{dressler87} and  elliptical galaxies of clusters 
at $\emph{z} =0.37$  \citep{ziegler97}. As can be seen in this Figure, 
the galaxies ID\,180, ID\,266, ID\,428 and  ID\,440 
follow the Faber-Jackson relation indicating that they  are early-type galaxies. 
The isophote maps of the first
three  galaxies are  typical of  ellipticals. In the case of 
ID\, 440 the isophotal map shows an extended structure probably due to the presence of a disc, 
therefore this galaxy may be  lenticular. Deep imaging  with higher spatial resolution is necessary to 
confirm the morphology of these galaxies.

The galaxy ID\,146 follows the Faber-Jackson relation marginally, that is, a
difference of about  2 mag  is present between this galaxy and the derived fit by \citep{ziegler97}
for Coma galaxies. Also, it can be seen in Figure 2 that 
the more external isophote shows an extended structure.  
Therefore, this galaxy probably has a morphology intermediate between  
lenticular and early spiral. The remaining galaxies ID\,120, ID\,237 and ID\,461,  have 
velocity dispersions of the order of the instrumental resolution and due to
their isophotal map morphology they might be spiral galaxies. 

The most important result obtained in this section is that the early-type field galaxies 
(ID\,146, ID\,180, ID\,266, ID\,428 and  ID\,440) have 
velocity dispersions and  luminosities comparable to those of  members of Coma and Virgo clusters 
and do not present a brightening of blue magnitude as observed by \citet{ziegler97} 
for elliptical galaxies at  $\emph{z} =0.37$. 


\section{Stellar population synthesis}
\label{sintesemodel}
We have performed the stellar population synthesis 
method developed by \citet{bica88}, in order to describe the age distribution of the stellar
population of these galaxies and get some insights about their metallicity.
This synthesis method uses the equivalent widths $W_{\lambda}$ of spectral absorption features
and measured continuum fluxes $F_{\lambda}$  at a given wavelength and compares them to the 
same quantities measured from a base of simple stellar population elements, which in turn 
have known ages and metallicities. The algorithm used is based on a upgraded version  of \citet{schmitt96}, 
which also corrects the effect of the internal extinction that affects the stellar population. 

The employed $W_{\lambda}$ were those defined by the Lick system \citep{worthey97,trager98}. 
Previous to the measurements of the $W_{\lambda}$ and $F_{\lambda}$
the spectra have been normalized at 5870 \AA\, for ID\,120 and
at 4200 \AA\, for ID\,146 and ID\,180.
The  $W_{\lambda}$ values for each galaxy 
are listed in Table \ref{lick}. The base elements that we have used were taken from the 
\citet{bruzual03} evolutionary 
stellar population models, which are based on a high resolution library of observed stellar 
spectra. This library allows us to derive a detailed spectral evolution of  simple 
stellar populations across the wavelength range of  3\,200 to 9\,500 $\AA$ with a wide range of 
metallicities. We have used the {\it Padova 1994 tracks} as recommended by \citet{bruzual03}.

\begin{figure}
\centering
\includegraphics*[width=\columnwidth]{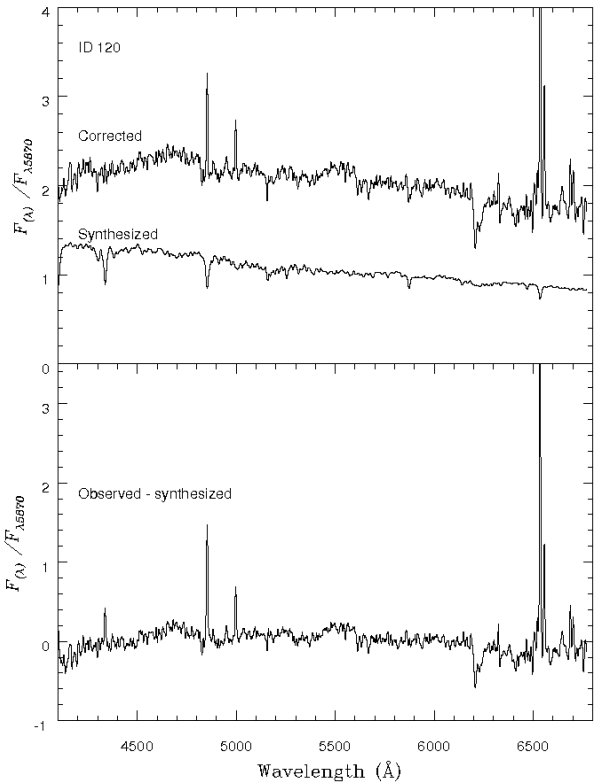}						     
\caption{Stellar population synthesis for ID\,120.
Top  panel: spectrum corrected for reddening and the synthesized spectrum. Bottom panel:
pure emission spectrum. The corrected spectrum has been shifted by a constant.}
\label{sint_120}
\end{figure}

\begin{figure}
\centering
\includegraphics*[width=\columnwidth]{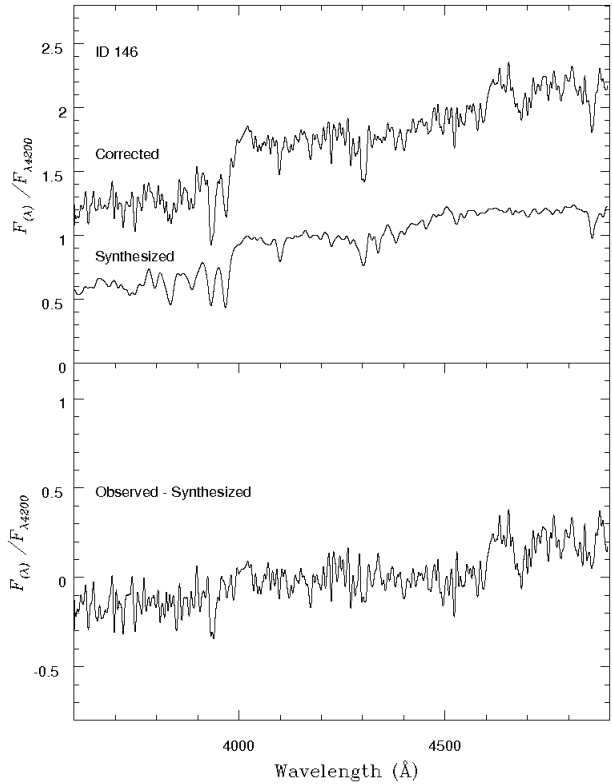}						     
\caption{Stellar population synthesis for ID\,146.
Top panel: spectrum corrected for reddening and synthesized spectrum. Bottom panel:
spectrum without the stellar component
The corrected spectrum has been shifted by a constant.}
\label{sint146}
\end{figure}

In the present work, we derived the stellar population  for ID\,120 
($\emph{z}=0.22$)
and ID\,146  ($\emph{z}=0.39$), the only galaxies in our sample with 
enough measured Lick indices that allow us to perform a reliable 
stellar population synthesis. 
Following the cosmological model adopted, the age of the universe at redshift $\emph{z}=0.22$ 
and $\emph{z}=0.39$ is around 10 and 9 Gyrs, respectively. Ten and 9 Gyrs  were the
upper limit ages for the spectral base of ID\,120 and ID\,146, respectively. 
The synthetic base consists of a set of 12 spectra 
with ages of 0.1, 1.0, 5.0, 9.0 and 10.0 Gyrs and metallicities of $Z=0.05$, 
$Z=0.02$ (solar) and $Z=0.004$. Note that we have chosen this range of metallicities because it was only possible to 
measure  a limited number of  Lick
indices in our spectra. The synthesis results for ID\,120 and ID\,146 are presented in 
Table \ref{sintese} as the percentual contribution of each template to 
the normalized spectrum, and the $E(B-V)$ values obtained from the synthesis.  
Figures \ref{sint_120} and \ref{sint146} show the spectra corrected for reddening and 
the synthesized population
spectra constructed  by the sum of the base spectra according to the proportions given 
by the synthesis. According to the stellar population synthesis, the galaxy ID\,120 has an 
important contribution of young stellar populations (0.1 Gyr), about 47\%. 
Despite this contribution  of a younger population, 
it is still made up of about 32 \% of an old stellar population (10 Gyr) 
and  the remaining is of intermediate age stellar population.  A  very large reddening 
of $E(B-V)=0.82$ was obtained from the synthesis. 
A solar metallicity is a better fit for this galaxy. In a general way, the estimates of old,
intermediate and young stellar populations for ID\,120 are in agreement with those 
derived by \citet{serote05} for field galaxies at intermediate redshifts.

For the galaxy ID\,146 its spectrum 
is best fitted by the template set of subsolar metallicity ($Z=0.004$) with
a dominant contribution of almost 60 \% of an old stellar population (10 Gyr) 
and a significant contribution of about  19\% and  17\%  
from intermediate components of 1 Gyr and 5 Gyr, respectively. Also,  
a young (0.1 Gyr) stellar population was marginally detected ($\approx$ 5 \%). 
The synthesis indicates a noticeable 
reddening in the spectrum, $E(B-V)=0.90$.

Differences between the stellar population 
of field galaxies and cluster members 
have been reported by \citet{serote05}. These authors
have performed stellar population synthesis 
in the cluster members of CL 0048-2942 ($\emph{z}\sim 0.64$), as well 
as, in the field galaxies ($ 0.2229 < \emph{z} < 0.6287$ and $ 0.6502 <
\emph{z} < 0.8215$) and have found in a general way, that field
galaxies seem to host less evolved stellar populations than cluster members. Unfortunately, 
since we can only derive the stellar population for two galaxies, ID\,120 and ID\,146, 
we cannot compare the stellar populations of field galaxies and 
cluster LCDCS-S001 members.

\begin{table}
\caption{Lick indices}
\label{lick}
\begin{tabular}{lccc}
\noalign{\smallskip}
\hline
\hline
\noalign{\smallskip}
    Equivalent width (\AA)                  & ID\,120         &ID\,146          & ID\,180\\
\noalign{\smallskip}
\hline
\noalign{\smallskip}  
H$\delta$                   &   ...          & 2.56$\pm$0.13  & ...     \\
\ion{Ca}{i} $\lambda$4227   &	...          & 1.57$\pm$0.14  & ...     \\
G $\lambda 4300$            &1.324$\pm0.13$  & 5.64$\pm$1.13  & 6.67$\pm$0.67 \\
$\rm H\gamma$ $\lambda 4340$&	...          & ...            & 2.91$\pm$0.87        \\
$\rm H\beta$ $\lambda 4861$ &	...          &  4.20$\pm$0.50 & ...     \\
\ion{Mg}{i} $\lambda 5167$  & 	1.70$\pm0.43$&  ...           & ...     \\
 NaD $\lambda 5890$         &	2.76$\pm0.44$&  10.14$\pm$0.61& ...     \\
\noalign{\smallskip}
\hline
\noalign{\smallskip}
\end{tabular}
\end{table}

\begin{table}
\caption{Stellar population synthesis results}
\label{sintese}
\begin{tabular}{lcccc}
\noalign{\smallskip}
\hline
\hline
\noalign{\smallskip}
\multicolumn{2}{c}{ID\,120}  & & \multicolumn{2}{c}{ID\,146}  \\
\noalign{\smallskip}
\cline{1-2}
\cline{4-5}
\noalign{\smallskip}
   Age (Gyr) &  $\lambda$ 5870 flux fraction  & & Age (Gyr)  & $\lambda$ 4200 flux fraction  \\      
\multicolumn{1}{c}{Z=0.02} &  (\%) & & \multicolumn{1}{c}{Z=0.004} & (\%)\\                                                
\noalign{\smallskip}
\hline
\noalign{\smallskip}  
0.1      &  46.64$\pm$2.91    & &   0.1     & 5.21$\pm$4.00     \\
1        &  8.51$\pm$5.63     & &   1       & 18.89$\pm$4.88 \\
5        &  13.56$\pm$9.10    & &   5       & 16.74$\pm$3.36 \\
10       &  32.29$\pm$6.27    & &   9       & 59.16$\pm$2.91 \\
$E(B-V)$ &   0.82             & &           $E(B-V)$    &0.90     \\
\noalign{\smallskip}
\hline
\noalign{\smallskip}
\end{tabular}
\end{table}


\section{Ionized gas}

As discussed earlier, the spectrum of the galaxy
ID\,120 presents conspicuous emission lines. 
In this section we investigate 
the physical properties and the ionization source of the gas in this galaxy.

First, we  subtracted the
contribution of the stellar population from the spectrum using the results of 
the stellar population synthesis. Then, we  estimated the 
line intensities  using Gaussian line profile fitting procedures, 
which are listed in Table \ref{intens}.

In order to identify the main excitation source of the gas in ID\,120 we  used the standard   
diagnostic diagrams with 
emission line ratios of easily observed lines, such as, 
the classical diagrams log ([\ion{O}{iii}]$ \lambda\,5007/\rm H \beta$) versus 
log ([\ion{N}{ii}]$ \lambda\,6584/\rm H \alpha$) and 
log ([\ion{O}{iii}]$ \lambda\,5007/\rm H \beta$) versus 
log ([\ion{S}{ii}]$ \lambda\,\lambda\,6717,6731/\rm H \alpha$) \citep{veilleux87}. 
They are applied either for nearby galaxies 
\citep{pastoriza99} and to galaxies at higher redshifts \citep{maier05,lamareille06a}.
In these diagnostic diagrams, the emission line ratios of ID\,120 are located  
\ion{H}{ii} region zone. 

We derived the electron density $N_{{\rm e}}$ from the [\ion{S}{ii}]$\lambda\,6716/\lambda\,6731$
intensity ratio by solving numerically the equilibrium equations for a $n$-level atom using 
the $temden$ routine of the $nebular$
package of  $STSDAS/IRAF$ assuming an electron temperature of 10\,000 K. 
The energy levels, transition probabilities and  
collisional strength values for  [\ion{S}{ii}] were 
taken respectively from \citet{bowen60}, \citet{keenan93} and \citet{ramsbottom96}. 
A value of $N_{{\rm e}}= 377\pm 184\, \mathrm{cm^{-3}} $ 
was estimated, which  is compatible  with the ones
derived in the range of $N_{{\rm e}} \approx 30 - 400\, \mathrm{cm^{-3}}$ by 
\citet{puech06} in 6 galaxies  at $\emph{z}= 0.55$   and 
are typical values of classical galactic \ion{H}{ii} regions \citep{copetti00} as well as of giant 
extragalactic \ion{H}{ii} regions \citep{castaneda92}.

Considering that the emission lines of ID\,120 are produced in an \ion{H}{ii} region
we have employed  the  photoionization code Cloudy/96.03 \citep{ferland02}  to produce a 
model of \ion{H}{ii} region that reproduces the observed 
emission lines intensities. 
Basically, the input nebular parameters in the models are the metallicity $Z$, ionization 
parameter $U$, upper stellar mass limit $M_{\rm up}$, age $A$ and electron density $N_{{\rm e}}$. 
The $M_{\rm up}$, $A$ and ${\log}(U)$
were chosen to be initially 40 M$_{\sun}$, 2 Myr and  -2.5, respectively, and 
were changed interactively until a suitable solution was found, following the 
same  fitting procedures adopted by \citet{dors06} (see description details in this paper). 
Besides, we have assumed in our models an electron density of  $N_{{\rm e}}= 377\, \mathrm{cm^{-3}}$
and  solar metallicity - since the synthesis results indicate that the spectrum of ID\,120  
is better fitted with this metallicity. 
  
We found an excellent agreement between the observational data
and the model of $12 + \log(\rm O/H)= 8.65$,   $M_{\rm up}$ = 100 $M_{\sun}$, and age of
5.5 Myr (see Table \ref{intens}). The oxygen abundance derived is compatible with the solar composition 
of $12 + \log(\rm O/H)= 8.69$
estimated by \citet{allende01} and with  measurements of abundance 
derived for other star forming galaxies at intermediate 
redshift \citep{barrientos96,contini02,lilly03,lamareille06b,mouhcine06}. For example,
\citet{mouhcine06} derived oxygen abundances for a sample of 40 luminous
field galaxies with redshifts in the range of $0.22 < \emph{z} < 0.8$ and found that their
$12 + \log(\rm O/H)$ values are ranging from 8.4 to 9.0, with a median of 8.7. 
Besides, these studies about metallicity in galaxies at intermediate redshift have
confirmed that the relation between luminosity and metallicity found in these galaxies 
is very similar to the relation obtained for galaxies in
the local universe. The values of $12 + \log(\rm O/H)= 8.65$ 
and M$_{B}= -18.75$  derived for ID\, 120 are compatible with the values of luminosity and 
metallicity for galaxies at intermediate redshifts found by these authors.

\begin{table}
\caption{Predicted and observed relative line fluxes ($\rm H\beta=100$) }
\centering
\label{intens}
\begin{tabular}{lcc}
\noalign{\smallskip}
\hline
\hline 
\noalign{\smallskip}
Line & Obs. & Mod. \\
\noalign{\smallskip}
\hline
[\ion{O}{iii}] $\lambda\,5007$ & 41.95$\pm$1.97 & 48.00 \\ 
$$[\ion{N}{ii}] $\lambda\,6584$& 98.50$\pm$4.24 & 98.00 \\
$$[\ion{S}{ii}] $\lambda\,6717$& 36.18$\pm$2.02 & 36.00 \\
$$[\ion{S}{ii}] $\lambda\,6731$& 32.61$\pm$3.61 & 33.00 \\
\noalign{\smallskip}
\hline
\noalign{\smallskip}
\end{tabular}
\end{table}
\section{Conclusions and final remarks}
We present spectroscopic and photometric analysis for eight field galaxies in the direction 
of the galaxy cluster LCDCS-S001. The photometric and spectroscopic 
data were obtained with the GMOS instrument in the Gemini South Observatory and consists of an 
$i'$-band image for the cluster field and spectra centered at 7500 \AA.
The main findings are the following:

\begin{enumerate}
\item
Redshifts of $\emph{z}=0.7464$  and $\emph{z}=0.7465$ were determined 
for the field galaxies ID\,440 and ID\,461, respectively. 
For the other six galaxies we have 
confirmed the redshift calculated by \citet{rembold06}. 
The redshifts of the field galaxies are in the range of $0.2201 < \emph{z} < 0.7784$
\item  
Stellar velocity dispersions were measured for 5 field galaxies and were found to be
in the range  of $200 < \sigma <  346\, \rm{ km\, s^{-1}}$.
\item
Blue luminosities were calculated for 6 galaxies which are  brighter than 
M$_{B}=-18.64$.
\item  
The galaxies ID\,180, ID\,266, ID\,428 and  ID\,461 
follow the Faber-Jackson relation and therefore are early-type. 
In these galaxies we did not detect a brightening of the $B$ luminosity with respect to 
the local galaxies.
\item
Lick indices were measured for  ID\,120 and ID\,146 and used to determine their  stellar 
population properties, by means of  spectral synthesis. 
In a general way, we found that
ID\,146  has an dominant contribution of an 
old stellar population ($\approx$ 10 Gyr) of subsolar metallicity.  
ID\,120  presents a spectrum with both emission and absorption lines, having an important 
contribution, of about 47 \%,  of  young stellar population and a significant contribution of 
about 32\% of older population 
($\approx$ 10 Gyr). The spectrum of ID\, 120 is better fitted with  solar metallicity. 
Large reddening values of   
$E(B-V)=0.90$  and $E(B-V)=0.82$ were derived for ID\,120 and ID\,146, respectively. 
\item
According to the classical diagnostic diagrams the emission lines present in 
$ \rm ID\, 120's$ spectrum  are produced in an
\ion{H}{ii} region. An electron density of $N_{{\rm e}}= 377\pm 184\, \mathrm{cm^{-3}} $
 was derived from the  [\ion{S}{ii}]$\lambda\,6716/\lambda\,6731$ ratio.
\item  
The values  of $12 + \log(\rm O/H)= 8.65$ and  M$_{B}=-18.75$
derived for ID\,120 are compatible with the values of luminosity and 
metallicity found by other authors 
for galaxies at intermediate redshifts.
\end{enumerate}

\thanks{Acknowledgments} 
We thank the referee, Dr. Florence Durret, for helpful comments and suggestions. 
This work has been partially supported by the Brazilian institution CNPQ.

\bsp

\label{lastpage}
\end{document}